\def\w{\omega}
\def\w0{\omega_0}
\def\k0{k_0}

\def\bk{{\bf k}}
\def\bkp{{\bf k}'}
\def\Bkj{B_{\bk j}}
\def\Bkjd{B_{\bk j}^\dagger}
\def\Bkjp{B_{\bk ' j'}}
\def\Bkjpd{B_{\bk 'j'}^\dagger}
\def\fkjl{\bf{f_{\bk j}^\ell}}
\def\fkjil{f_{\bk j}^{i\ell}}
\def\fkjul{f_{\bk j}^{1\ell}}
\def\fkjdl{f_{\bk j}^{2\ell}}
\def\fkjpul{f_{\bk 'j'}^{1\ell}}
\def\fkjpdl{f_{\bk 'j'}^{2\ell}}
\def\fkjpil{f_{\bk 'j'}^{i\ell}}
\def\br{{\bf r}}
\def\ekj{\hat{e}_{\bk j}}
\def\bkj{b_{\bk j}}
\def\bkjd{b_{\bk j}^\dagger}
\def\ail{a_{i\ell}}
\def\aild{a_{i\ell}^\dagger}
\def\tkjl{\bf{t^{\ell}_{\bk j}}}
\def\tilkj{t^{i\ell}_{\bk j}}
\def\rkjl{\bf{r^{\ell}_{\bk j}}}
\def\rilkj{r^{i\ell}_{\bk j}}
\def\tulkj{t^{1\ell}_{\bk j}}
\def\rulkj{r^{1\ell}_{\bk j}}
\def\tdlkj{t^{2\ell}_{\bk j}}
\def\rdlkj{r^{2\ell}_{\bk j}}
\def\Tkjpkj{T_{\bk j}^{\bk'j'}}
\def\Rkjpkj{R_{\bk j}^{\bk'j'}}
\def\Ml{\mathbf{M}_{\ell}}
\def\Akjil{\lambda_{i\ell}^{\bk j }}
\def\Akjul{\lambda_{1\ell}^{\bk j }}
\def\Akpjpul{\lambda_{1\ell}^{\bkp j' }}
\def\Akjdl{\lambda_{2\ell}^{\bk j }}
\def\Akpjpdl{\lambda_{2\ell}^{\bkp j' }}
\def\Ckjil{\tau_{i\ell}^{\bk j }}
\def\etakjkjp{\eta_{\bk j}^{\bkp j'}}
\def\mukjkjp{\mu_{\bk j}^{\bkp j'}}

\documentclass[preprint,aps,showpacs]{revtex4}

\begin{document}

\title{An exactly solvable model of two three-dimensional harmonic oscillators interacting with the quantum electromagnetic field: the Casimir-Polder potential} 

\author{F. Ciccarello\mbox{${\ }^{*}$}, E. Karpov\mbox{${\ }^{**}$}, R. Passante\mbox{${\ }^{***}$}}
\affiliation{
\mbox{${\ }^{*}$} Dipartimento di Fisica e Tecnologie Relative 
dell'Universit\`{a} degli Studi di Palermo and INFM, Viale delle Scienze,
Edificio 18, 90128 Palermo, Italy \\
\mbox{${\ }^{**}$} Quantum Information and Communication (QUIC), 
CP 165/59, 
Ecole Polytechnique, Universit\'{e} Libre de Bruxelles, 
50 av. F. D. Roosevelt, B-1050 Bruxelles, Belgium  \\
\mbox{${\ }^{***}$}Dipartimento di Scienze Fisiche ed Astronomiche 
dell'Universit\`{a} degli Studi di Palermo and INFM, Via Archirafi 36, I-90123 Palermo, Italy
 }

\email{roberto.passante@fisica.unipa.it}


\begin{abstract}
We consider two three-dimensional isotropic harmonic oscillators interacting with
the quantum electromagnetic field in the Coulomb gauge and within dipole
approximation. Using a Bogoliubov-like transformation, we can obtain transformed
operators such that the Hamiltonian of the system, when expressed in terms of these operators, 
assumes a diagonal form. We are also able to obtain an expression for the energy shift
of the ground state, which is valid at all orders in the coupling constant. From this energy
shift the nonperturbative Casimir-Polder potential energy between the two oscillators can
be obtained. When approximated to the fourth order in the electric charge, the well-known
expression of the Casimir-Polder potential in terms of the polarizabilities of the oscillators is recovered.
\end{abstract}

\maketitle

\section{\label{sec:1}Introduction}

The Casimir-Polder potential is a long range interaction between neutral
atoms/molecules in vacuo \cite{CP48}. This potential arises from the interaction of
the two atoms with the common radiation field. It is also a result of the
quantum field fluctuations in the vacuum state, and it is one of the few
unambiguous manifestations of the quantum nature of the electromagnetic
field \cite{CPP95}.

The Casimir-Polder potential is usually obtained by fourth-order perturbation
theory \cite{CT98}. Higher-order corrections exist, as well as 	non-additive components
in the case of three or more atoms \cite{PT85}. 
More recently Casimir-Polder forces for excited atoms have been considered \cite{PT93},
even if the precise meaning of a dressed excited state is not clear \cite{PPT91,PPR00,KOPP02}, 
as well as time-dependent Casimir-Polder energies for partially dressed atoms \cite{PP03}
or atoms in an excited-state \cite{RPP04}.

The Casimir-Polder potential can be expressed in terms of the dynamical polarizabilities of 
the two atoms/molecules, and this makes clear that a similar  potential is expected whenever
we consider two neutral systems with a discrete spectrum interacting with a quantum field (not
necessarily in the framework of quantum electrodynamics, but also in condensed matter physics,
for example).

In this paper we consider two harmonic oscillators interacting with the electromagnetic radiation field.
This system has the advantage that exact solutions can be obtained by a
Bogoliubov-like transformation, as we will show in this paper. 
It is also a good representation
for atomic systems; approximating atoms by harmonic oscillators is a common procedure, due
to the resulting simplification of calculations (for a few examples, see \cite{Senitzky98,KR96,RZ76}).  
An harmonic oscillator interacting with a one-dimensional scalar field
has been recently considered in order to study phenomena relevant to quantum optics,
and exactly solved by a Bogoliubov transformation \cite{KPTP00,KOPP02}; similar methods have 
been used also for the description of unstable states in relativistic field theory \cite{AGP01}.
An exact solution for two two-level atoms interacting with a one-dimensional
scalar field, in the one excitation sector, has been recently obtained for the description of 
decaying states in the rotating wave approximation \cite{OK04}. The rotating wave approximation, 
however, prevents application of these results to the Casimir-Polder potential, 
where the virtual processes play an essential role.

We are mainly interested in the study of the Casimir-Polder potential. We therefore consider
two spatially separated isotropic three-dimensional harmonic oscillators, 
(thus having a spherical symmetry), interacting with the quantum electromagnetic field.
Counterrotating terms and the related virtual processes are included in the Hamiltonian.
By a Bogoliubov-like transformation we can obtain an exact diagonalization of the
Hamiltonian describing this system. 
The Hamiltonian of the system, in terms of the transformed operators, has the form
of a free field Hamiltonian.
Moreover, using the inverse transformations we are also able to obtain the exact energy shift
of the ground state of the system, due to the interaction between the oscillators
and the radiation field. This result is exact
within the model considered and valid at all orders in the coupling constant. The part of
the energy shift depending on the distance between the two oscillators yields the
Casimir-Polder potential between the ground-state oscillators, 
both in the so-called near and far zones. 
Known results are reproduced by approximation to the fourth-order
in the electric charge. 
Due to the generality of our model, and the fact that 
the validity of our results is not limited by perturbation theory, the results obtained in this paper can be also
used to obtain Casimir-Polder-like potentials in all cases in which the coupling constant
is not small.
We also expect that our nonperturbative approach can be useful in dealing with
Casimir-Polder forces for atoms in excited states, being able to overcome the conceptual
difficulties related to the vanishing energy denominators occurring in the perturbative
expansions when excited states are considered \cite{PPT91, OPP01}. 

In Section \ref{sec:2} we introduce the Hamiltonian describing our system and  the 
transformation used to obtain the new operators; in terms of the new operators, the Hamiltonian
assumes a diagonal form.
The explicit relation between the
{\it new} and {\it old} operators is also obtained, as well as the inverse relations. 
In section \ref{sec:3} we obtain the energy
shift of the ground state of the interacting system, from which the Casimir-Polder potential
between the two isotropic harmonic oscillators is obtained. The physical meaning of our results
is discussed, as well as possible applications to Casimir-Polder forces for excited atoms.

\section{\label{sec:2}Hamiltonian and Bogoliubov transformation}

We consider two isotropic three-dimensional harmonic oscillators with frequency $\w0 = c\k0$
interacting with the electromagnetic field in the Coulomb gauge. In the
multipolar coupling scheme the Hamiltonian of this system is \cite{CPP95}
\begin{equation}
H = \sum_{i=1,2}\sum_{\ell =x,y,z} \k0 a_{i\ell}^\dagger a_{i\ell} +
\sum_{\bk j} k \Bkjd \Bkj + H_{int}^1+ H_{int}^2
\label{eq:1}
\end{equation}
(we use units such that $\hbar =1$ and $c=1$).
The index $i$ labels the two oscillators ($i=1,2$), the index $\ell$ the space
direction of the oscillator ($\ell = x,y,z$), $a_{i\ell}^\dagger , a_{i\ell}$ are the
creations and annihilation operators of the two three-dimensional oscillators and
$\Bkjd , \Bkj$ for the field ($j$ is the polarization index), $k = \mid \bk \mid$.
The creation and annihilation operators in (\ref{eq:1}) satisfy the usual
bosonic commutation rules
\begin{equation}
\left[ a_{i\ell}, a_{i'\ell '}^\dagger \right] = \delta_{ii'} \delta_{\ell \ell '} \; , \; \;
\left[ \Bkj ,  \Bkjpd \right] = \delta_{\bk \bkp} \delta_{jj'}.
\label{eq:1a}
\end{equation}
$H_{int}^1$ and $H_{int}^2$ are the interaction Hamiltonians of the
oscillators 1 and 2 with the field, respectively. Within the dipole approximation, they are given by
\begin{equation}
H_{int}^1 = \sum_{\ell =x,y,z} \sum_{\bk j} \fkjul
\left( a_{1\ell} + a_{1\ell}^\dagger \right) 
\left( \Bkj - \Bkjd \right)
\label{eq:2}
\end{equation}

\begin{equation}
H_{int}^2 = \sum_{\ell =x,y,z} \sum_{\bk j} \fkjdl
\left( a_{2\ell} + a_{2\ell}^\dagger \right) 
\left( \Bkj e^{i\bk \cdot \br} - \Bkjd e^{-i\bk \cdot \br} \right)
\label{eq:3}
\end{equation}
where the atom 1 is supposed to be located at the origin
and the atom 2 at $\br$, and the coupling constants are

\begin{equation}
f_{\bk j}^{i\ell} = -i \sqrt{\frac{2\pi k}{V}} \mbox{\boldmath $\mu$}_{i\ell}
\cdot \ekj 
\end{equation}
$\mbox{\boldmath $\mu$}_{i\ell}$ is the transition dipole moment
of the atom $i$ in the $\ell$ direction and $V$ is the quantization volume;
the dipole moments are supposed real.

In order to diagonalize the Hamiltonian (\ref{eq:1}),
we look for a linear transformation of the operators, similarly
to the case of just one one-dimensional oscillator interacting
with a scalar field \cite{KPTP00}. Thus we write 
the new operators $\bkj , \bkjd$
in terms of the old operators in the following form
\begin{equation}\label{sviluppobB} 
\bkjd = \sum_{i=1,2} \sum_{\ell = x,y,z} 
\left( \tilkj \aild + \rilkj \ail \right) +
\sum_{\bk' j'}
\left( \Tkjpkj \Bkjpd + \Rkjpkj \Bkjp \right)
\label{eq:4}
\end{equation}
and its Hermitian conjugate for the annihilation operators $\bkj$.
The coefficients $\tilkj , \rilkj , \Tkjpkj ,  \Rkjpkj$ are chosen in such a
way that the ``new" operators satisfy the {\it free-field} commutation 
relations
\begin{eqnarray}
\left[ H, \bkjd \right] &=& k \: \bkjd
\nonumber \\
\left[ H, \bkj \right] &=& -k \: \bkj
\label{eq:5}
\end{eqnarray}

Substitution of (\ref{eq:4}) into (\ref{eq:5}) yields the following set of 
coupled equations for the coefficients

\begin{eqnarray}
(k -\k0 ) \tulkj &=& \sum_{\bk' j'} \fkjpul 
\left( \Tkjpkj + \Rkjpkj \right)
\label{eq:6a} \\
(k +\k0 ) \rulkj &=& \sum_{\bk' j'} \fkjpul 
\left( \Tkjpkj + \Rkjpkj \right)
\label{eq:6b} \\
(k -\k0 ) \tdlkj &=& \sum_{\bk' j'} \fkjpdl 
\left( \Tkjpkj e^{i\bk '\cdot \br} + \Rkjpkj e^{-i\bk '\cdot \br}  \right)
\label{eq:6c} \\
(k +\k0 ) \rdlkj &=& \sum_{\bk' j'} \fkjpdl 
\left( \Tkjpkj e^{i\bk '\cdot \br} + \Rkjpkj e^{-i\bk '\cdot \br}  \right)
\label{eq:6d} \\
(k-k') \Tkjpkj &=& \sum_\ell \fkjpul \left( \rulkj -\tulkj \right) +
\sum_\ell \fkjpdl e^{-i \bk '\cdot \br} 
\left( \rdlkj - \tdlkj \right)
\label{eq:6e} \\
-(k+k') \Rkjpkj &=& \sum_\ell \fkjpul \left( \rulkj -\tulkj \right) +
\sum_\ell \fkjpdl e^{i \bk '\cdot \br} 
\left( \rdlkj - \tdlkj \right)
\label{eq:6f}
\end{eqnarray}
This set of coupled equations can be solved exactly. The solution can be expressed in terms of the resolvent
\begin{equation}
\left( G_{i}(k) \right)^{-1} = \k0^2- k^2 -2\k0 \sum_{\bkp j'} \frac {2k'}{k^2-k'^2}
\fkjpil \fkjpil
\label{eq:8}
\end{equation}
and the r-dependent function
\begin{equation} 
\sigma_\ell (k,r) = -2\k0 \sum_{\bkp j'} 
\left( \frac {e^{i\bkp \cdot \br}}{k+k'} - \frac {e^{-i\bkp \cdot \br}}{k-k'} \right)
\fkjpul \fkjpdl
\label{eq:9}
\end{equation}

For isotropic oscillators, as we assume, the resolvent (\ref{eq:8})
does not depend on $\ell$. Explicit expressions of (\ref{eq:8})
and (\ref{eq:9}) in the continuous limit are given in the Appendix \ref{AppendixA}.
In this limit, the two functions (\ref{eq:8}) and (\ref{eq:9}) have poles for $k=k'$,
and must be extended to the complex plane
\begin{eqnarray}
G^{(\pm)}_i(k) &=& G_i(k \pm i\epsilon) 
\label{eq:8a} \\
\sigma^{(\pm)}_\ell (k,r) &=& \sigma_\ell (k \pm i\epsilon ,r) 
\label{eq:9a}
\end{eqnarray}
However, whenever unnecessary, for simplicity of notations we shall omit the
apix relative to the choice of the analytic continuation of these functions.
For convenience of notations the two functions defined in
(\ref{eq:8}) and (\ref{eq:9})  can be expressed as the elements of
a symmetric 2x2 matrix, 
with $ \left( G_1 \right)^{-1}$ and $\left( G_2 \right)^{-1}$ as 
diagonal elements and $\sigma_{\ell}$ as off-diagonal elements,
apart from a proportional factor. So we define the matrix
\begin{equation}
\Ml (k,r) = \frac {G_1 G_2}{1-\sigma_\ell^2 G_1 G_2}
\pmatrix{G_2^{-1} & -\sigma_\ell \cr
	-\sigma_\ell  & G_1^{-1} \cr} 
\label{eq:11}
\end{equation}
We also define the following two-dimensional column
vectors
\begin{equation}
\fkjl = 
\pmatrix{ \fkjul \cr \fkjdl e^{i\bk \cdot \br} \cr}
\label{eq:11a}
\end{equation}

\begin{equation}
\tkjl = 
\pmatrix{ \tulkj \cr \tdlkj \cr} ; \hspace{20pt}
\rkjl = 
\pmatrix{ \rulkj \cr \rdlkj \cr}
\label{eq:11b}
\end{equation}

The following relations exist among $\rkjl$ and $\tkjl$,
and among $\Rkjpkj$ and $\Tkjpkj$ 
\begin{eqnarray}
\rkjl  &=& \frac{k-\k0}{k+\k0} \, \tkjl
\label{eq:11c} \\
\Rkjpkj &=& \frac{k'-k}{k'+k} \, T_{\bk j}^{-\bkp j'}
\label{eq:11d} 
\end{eqnarray}

The solution for the coefficients defined in (\ref{eq:4}) can
be expressed in the compact form
\begin{eqnarray} 
\tkjl &=& - (k +\k0 ) \Ml \, \fkjl 
\label{eq:15a} \\
\rkjl &=&  -\left( k - \k0 \right) \Ml \, \fkjl
\label{eq:15b} \\
\Tkjpkj &=& \delta_{jj'} \delta_{\bk \bkp} 
+ \frac {2\k0}{k -k'} \left( {\bf f}_{\bkp j'}^\ell \right)^\dagger \Ml \, \fkjl
\label{eq:15c} \\
\Rkjpkj &=&  -\frac {2\k0}{k' +k}  \left( {\bf f}_{\bkp j'}^\ell \right)^\dagger \Ml \, \fkjl
\label{eq:15d}
\end{eqnarray}

The correct commutation relation for the transformed operators
\begin{equation}
\left[ \bkj , b_{\bkp j'}^\dagger \right] = \delta_{jj'} \delta_{\bk \bkp}
\label{eq:commrel}
\end{equation}
can be proved by expressing $\bkj$ and $b_{\bkp j'}^\dagger$ in terms of the ``old" operators by means of 
eq. (\ref{sviluppobB}) and eqs. (\ref{eq:15a},\ref{eq:15b},\ref{eq:15c},\ref{eq:15d}). 
Equations (\ref{usefulformulaF1u}) and (\ref{usefulformulaF1d}) of Appendix \ref{AppendixA} have been used 
in order to obtain the commutation relations (\ref{eq:commrel}).

The inverse relations, expressing old operators in terms of new
operators, can be also obtained in the form
\begin{eqnarray}
\aild &=& \sum_{\bk j} \left( \Akjil \, \bkjd 
+ \Ckjil \, \bkjd \right)
\label{eq:10a} \\
\Bkjd &=& \sum_{\bkp j'} \left( \mukjkjp\, b_{\bkp j'}^\dagger
+ \etakjkjp\,  b_{\bkp j'} \right)
\label{eq:10b}
\end{eqnarray}
and the Hermitian conjugate relations for $\ail$ and $\Bkj$. The coefficients
appearing in eqs. (\ref{eq:10a},\ref{eq:10b}) can be obtained with the same procedure
used for the ``direct'' relations. The result for the $\lambda$ coefficients is
\begin{eqnarray}
\Akjul &=& \frac {(k+\k0)\,G_1(k)}{1-\sigma_\ell^2(k,r)G_1(k) G_2(k)}
\left( \fkjul - \fkjdl e^{-i\bk \cdot \br} \sigma_\ell (k,r) G_2(k) \right)
\label{eq:12a} \\
\Akjdl &=& \frac {(k+\k0)\,G_2(k)}{1-\sigma_\ell^2(k,r)G_1(k) G_2(k)}
\left( \fkjdl e^{-i\bk \cdot \br} - \fkjul \sigma_\ell (k,r) G_1(k) \right)
\label{eq:12b}
\end{eqnarray}
The other coefficients may be expressed in terms of the $\lambda$ coefficients as
\begin{eqnarray}
\Ckjil &=& -\frac {k-\k0}{k+\k0} \, \left( {\Akjil} \right)^\star
\label{eq:12c} \\
\mukjkjp &=& \delta_{jj'} \delta^{3}_{\bk \bkp}
+\frac {2\k0}{(k' + \k0 )(k'-k)} 
\left( \sum_\ell \fkjul \Akpjpul +\sum_\ell e^{i\bk \cdot \br}
\fkjdl \Akpjpdl \right) 
\label{eq:12d} \\
\etakjkjp &=& \frac {k'-k}{k'+k} \left( {\mu^{\bkp j'}_{-\bk j}} \right)^\star
\label{eq:12e}
\end{eqnarray}

\section{\label{sec:3}The ground-state energy}

We now evaluate the energy shift of the ground-state energy due to the
interaction of the two harmonic oscillators with the electromagnetic field.
The ground-state energy has terms which depend on the distance $r$ between the two
oscillators; this yields a Casimir-Polder potential between the two
oscillators.

In order to evaluate the energy of the ground state we substitute the inverse
relations (\ref{eq:12a},\ref{eq:12b},\ref{eq:12c},\ref{eq:12d},\ref{eq:12e}) in the 
Hamiltonian (\ref{eq:1}). After this substitution, the Hamiltonian in terms of
the new operators assumes a diagonal form 
\begin{equation}
H = \sum_{\bk j} k \ \bkjd \bkj + E_0
\label{eq:13a}
\end{equation}
where $E_0$ is the ground-state energy.
The part of $E_0$ depending from the interatomic distance $r$ gives the
Casimir-Polder potential energy between the oscillators. 
After extensive algebraic calculations, and use of 
eqs. (\ref{usefulformulaF2u}) and (\ref{usefulformulaF2d}) of Appendix \ref{AppendixA},  
we obtain
\begin{eqnarray} 
E_0 &=& \k0 \sum_{\bk j}\left(3k^{2}-2k\k0-\k0^{2}\right)	 
\sum_{i\ell}\frac{\Akjil \left( {\Akjil} \right)^\star}{(k+\k0)^2} 
- 4 \omega_{0}^{2} \sum_{\bk j \bkp j'}\frac{k}{(k+k')^{2}} 
\nonumber \\
&\times& \sum_{\ell}\left[ (f_{\bk j}^{1\ell}) ^{2}\,\frac{\Akpjpul 
\left( {\Akpjpul} \right)^\star}{(k'+\k0)^{2}}+(f_{\bk j}^{2\ell}) ^{2}\,
\frac{\Akpjpdl \left( {\Akpjpdl} \right)^\star}{(k'+\k0)^{2}} \right.
\nonumber \\
&+& \left. \left(f_{\bk j}^{1\ell} f_{\bk j}^{2\ell}e^{-i\bk \cdot \br}
\frac{\Akpjpul \left( {\Akpjpdl} \right)^\star}{(k'+\k0)^{2}}+\,c.c.\right)\right]
\label{eq:13}
\end{eqnarray}
(it should be note that the $\lambda$ functions in (\ref{eq:13}) depend on $r$).
This expression is valid at any order in the electric charge.
In the continuous limit, using the results of Appendix \ref{AppendixA}, we can obtain an explicit 
expression of the shift of the ground state energy $E_{0}$. 
By retaining only the $r$ dependent parts of eq. (\ref{eq:13}) up to the $4^{th}$ order in the charge,
approximating to the far zone $(k,k' \ll \k0 )$ \cite{CPP95}, the following expression for the interatomic potential $V_{CP}(r)$
is obtained
\begin{equation} \label{VCP12} 
V_{CP}(r)=\frac{16\,\mu^{2}_{1}\mu^{2}_{2}}{\pi^2 \k0^{2}}\,(2I_{1}+I_{2})\frac{1}{r^{7}}	
\end{equation}
where $I_{1}$ and $I_{2}$ stand for the following integrals

\begin{equation} \label{I1} 
I_{1}=\int_{0}^{\infty}\! dx \, \int_{0}^{\infty} \! dy \, \frac{x^{2}y^{4}}{x^{2}-y^{2}}
\left[ x j_{0}(x)\left(j_{0}(y)-\frac{j_{1}(y)}{y} \right)+j_{1}(x)\left(\frac{3j_{1}(y)}{y}-j_{0}(y) \right) \right] 	
\end{equation}

\begin{equation} \label{I2} 
I_{2}=\int_{0}^{\infty} \! dx  \int_{0}^{\infty}\! dy \, \frac{x^{2}y^{4}}{(x+y)^{2}}
\left[ x j_{0}(x)\left(j_{0}(y)-\frac{j_{1}(y)}{y} \right)+j_{1}(x)\left(
\frac{3j_{1}(y)}y - j_{0}(y) \right)   \right]	
\end{equation}

Once the above integrals are computed by standard methods as $I_{1}=-\frac{23}{16 \pi}$ and $I_{2}=\frac{23}{16 \pi}$,  the final result is 
\begin{equation} \label{VCPfinale} 
V_{CP}(R)=-\frac{23}{4\pi}\frac{\alpha_{1}\alpha_{2}}{r^{7}}	
\label{eq:14}
\end{equation}
where $\alpha_{i}$ ($i=1,2$) is the static polarizability of the i-th three-dimensional isotropic harmonic oscillator, that is 
\begin{equation}
\alpha_{i}=\frac{2}{3}\sum_{\ell=x,y,z} \frac{\mu_{i\ell}^{2}}{\k0} =\frac{2\,\mu_{i}^{2}}{\k0}
\label{eq:15}
\end{equation}
The quantity (\ref{eq:14}) is the well-known expression of the Casimir-Polder potential in the far
zone and for isotropic molecules \cite{CPP95,CT98}. Therefore, in the appropriate limits, 
the exact result ({\ref{eq:13}) of the ground-state
energy $E_0$ yields the usual Casimir-Polder potential energy and this confirms the validity of
our approach.  

\section{\label{sec:4}Conclusions}

We have considered a system of two isotropic three-dimensional harmonic oscillators interacting with
the electromagnetic radiation field, described in the Coulomb gauge and in the dipole approximation. 
Through an appropriate
transformation we are able to diagonalize the Hamiltonian of this system and obtain exact solutions. 
The diagonalized Hamiltonian is an infinite set of harmonic oscillators with the same spectrum of the free electromagnetic field, plus an energy shift. We have obtained an expression of this energy shift valid at all orders in the coupling constant, which contains terms depending from the distance $r$ between the two three-dimensional oscillators. These $r$-dependent
terms yield a potential energy between the two oscillators
in the ground state of the interacting system, 
i.e. the Casimir-Polder interaction. Our results allow to obtain this energy exactly, at any order in the 
coupling constant. Approximation to the fourth order in the coupling constant reproduces known results of the Casimir-Polder long-range interaction in terms of the static polarizabilities of the oscillators (in the far zone). 
We plan to discuss in a future publication the extension of the results obtained in this paper to the case of the Casimir-Polder potential for excited states. In this case, we should be able to obtain the potential without the well-known difficulties related to vanishing energy denominators that appear when the stationary perturbation theory is applied to excited states.

Finally, we wish to stress that our method is quite general and not limited to systems  interacting with the electromagnetic field.
Our results can be easily extended to other matter-field interacting systems, for example the interaction with phonons. Thus the method presented in this paper could be particularly useful for various physical systems with a strong coupling constant.

\appendix

\section{\label{AppendixA}The functions $G_i(k)$ and $\sigma (k,r)$ in the continuous limit}

In this Appendix we derive explicit expressions for the functions defined in
(\ref{eq:8}) and (\ref{eq:9}), in the case of isotropic oscillators, as well as quantities useful for the explicit evaluation of the commutation relations (\ref{eq:commrel}) and of energy shift $E_{0}$ given in Sec. III. The hypotheses of isotropic 
oscillators allows significant simplifications of the final equations.
In equation (\ref{eq:8}) the summation over the polarization index $j'$ can be done using
$\sum_{j'} (\hat{e}_{\bkp j'})_m (\hat{e}_{\bkp j'})_n = \delta_{mn} -\hat{k'}_m \hat{k'}_n$.
Also, in the continuous limit
\begin{equation}
 \sum_{\bkp} \rightarrow \frac V{(2\pi )^3} \int dk' {k'}^2 \int d\Omega '
\label{eq:A1}
\end{equation}
Thus, after polarization sum and angular integration we obtain
\begin{equation}
\frac V{(2\pi )^3}
\int d\Omega ' \sum_{j'} f_{\bkp j'}^{i\ell} f_{\bkp j'}^{i\ell '}
= -\frac {2}{3\pi} k' \mu_i^{2} \delta_{\ell \ell '} 
\label{eq:A2}
\end{equation}
where we have used that for symmetrical oscillators the strength of the
dipole moment in the three spatial directions is the same, here indicated
with $\mu_i$ ($i=1,2$).

We can proceed similarly for the quantity appearing in (\ref{eq:9}). After summation over
polarizations and angular integration, we obtain
\begin{equation} 
\frac V{(2\pi )^3}
\int d\Omega ' e^{\pm i \bkp \cdot \br} \sum_{j'} f_{\bkp j'}^{1\ell} f_{\bkp j'}^{2\ell '}
= -\frac{k'}{\pi} h_{\ell}(k'r) \mu_1 \mu_2 \delta_{\ell \ell '} 
\label{eq:A3}
\end{equation}
The function $h_\ell (k'r)$ is defined by
\begin{equation}
h_\ell (k'r) = \left\{ 
\begin{array}{cc}
	 j_0(k'r) - \frac 1{k'r} j_1(k'r) & \mbox{for $\ell = x,y$} \\
	 \frac 2{k'r} j_1(k'r) & \mbox{for $\ell = z$}
\end{array}
\right.
\label{eq:A4}
\end{equation}
where the axis $z$ has been taken along the direction of $\br$, and
$j_0,j_1$ are spherical Bessel's functions.

Thus, in the case of isotropic oscillators the functions (\ref{eq:8}) and
(\ref{eq:9}) take the form
\begin{eqnarray}
\left( G_i (k) ) \right)^{-1} &=& \k0^2 - k^2 - \frac{4}{3\pi} \k0 \mu_i^2 
\int dk' \frac {2{k'}^4}{{k'}^2 - k^2}
\label{eq:A4a} \\
\sigma_\ell (k,r) &=& -\frac{2}{\pi} \k0  \mu_1 \mu_2 \int dk' \frac {2{k'}^4}{{k'}^2 - k^2}
h_\ell (k'r)
\label{eq:A4b}
\end{eqnarray}
Equation (\ref{eq:A4a}) contains an ultraviolet divergence that, however,
is inessential for our purposes. (\ref{eq:A4b}) is convergent due to the oscillating
behaviour of $h_\ell (k'r)$.

By using the definitions (\ref{eq:8}) and (\ref{eq:9}) of $\left( G_{i}(k) \right)^{-1}$ and $\sigma_{\ell}(k,r)$, the following identities,
which have been used to obtain the commutation relations (\ref{eq:commrel}) and to evaluate the energy shift $E_0$, hold
  
\begin{eqnarray} \label{usefulformulaF1u}
\sum_{\bk j}  (\fkjil )^{2}F_{k',\, \tilde{k}}^{(1)}(k) 
&=&\frac{1}{2\k0} \left[\frac{\left( G_{i}(k') \right)^{-1}-\left( G_{i}(\tilde{k}) \right)^{-1}}{k' - \tilde{k}} +
(k'+\tilde{k})\right] \\ 
\label{usefulformulaF1d}
\sum_{\bk j} \, \fkjul \fkjdl \, e^{\pm i\bk \cdot \br} 
F_{k',\, \tilde{k}}^{(1)}(k) &=&\frac{\sigma_{\ell}(k',r)-\sigma_{\ell}(\tilde{k},r)}{2\k0 (\tilde{k}-k')}
\end{eqnarray}

\begin{eqnarray} \label{usefulformulaF2u}
\sum_{\bk j} k \, (\fkjil )^{2}F_{k',\, \tilde{k}}^{(2)}(k) 
&=& \frac {\k0}2 - \frac{k'^3 + \tilde{k}^3 + k' \left( G_{i}(k') \right)^{-1} + \tilde{k} \left( G_{i}(\tilde{k}) \right)^{-1}}
{2\k0 (k' + \tilde{k})} \\
\label{usefulformulaF2d} \sum_{\bk j} k \, \fkjul \fkjdl \, e^{\pm i\bk \cdot \br} 
F_{k',\, \tilde{k}}^{(2)}(k) &=& \frac{1}{2\k0 (k'+\tilde{k})}\left[k'\sigma_{\ell}(k',r)+\tilde{k}\sigma_{\ell}(\tilde{k},r)\right]
\end{eqnarray}
where $F_{k',\, \tilde{k}}^{(1)}(k)$ and $F_{k',\, \tilde{k}}^{(2)}(k)$ are given by

\begin{eqnarray}
F_{k',\, \tilde{k}}^{(1)}(k) &=& \frac{1}{(k'-k) (\tilde{k}-k)}-\frac{1}{(\tilde{k}+k)(k'+k)}
=\frac{2k}{\tilde{k}-k'}\left(\frac{1}{k^2 - \tilde{k}^2} - \frac{1}{k^2 - k'^2}\right)  \\
F_{k',\, \tilde{k}}^{(2)}(k) &=& \frac{1}{(k'-k)(\tilde{k}+k)}+\frac{1}{(\tilde{k}-k)(k'+k)}
=-\frac{2}{\tilde{k}+k'}\left(\frac{k'}{k^{2}-k'^2} +\frac{\tilde{k}}{k^{2}-\tilde{k}^2}\right)
\end{eqnarray}

\begin{acknowledgments}
The authors wish to thank Dr. G. Ordonez for many discussions on the subject of this paper.
This work was in part supported by the bilateral Italian-Belgian project on 
``Casimir-Polder forces, Casimir effect and their fluctuations" 
and by the bilateral Italian-Japanese project 15C1 on ``Quantum
Information and Computation" of the Italian Ministry for Foreign Affairs. 
Partial support by Ministero dell'Universit\`{a} e della Ricerca Scientifica 
e Tecnologica and by Comitato Regionale di Ricerche Nucleari e di Struttura della Materia
is also acknowledged.
\end{acknowledgments}

\end{document}